\documentclass[superscriptaddress,twocolumn,floats,showpacs,prl,amsmath,amssymb,floatfix,balancelastpage]{revtex4}

\input epsf
\usepackage{psfig}
\usepackage{graphicx}

\newcommand{\bea}{\begin{eqnarray}}
\newcommand{\eea}{\end{eqnarray}}
\newcommand{\bean}{\begin{eqnarray*}}
\newcommand{\eean}{\end{eqnarray*}}

\begin{document}

\title{On the origin of the cosmic microwave background anisotropies}
\author{Ria Follop}
\affiliation{Institute of Fundamental and
  Outstanding Questions, Department of Cosmology and Metaphysics,
  Online University, Internet}
\author{Ana\"is Rassat}
\affiliation{Department of Physics and Astronomy, University College London,  London, WC1E 6BT}
\author{ Asantha Cooray}
\affiliation{Department of Physics and Astronomy, 4186 Frederick Reines Hall, University of California, Irvine, CA 92697}
\author{Filipe B. Abdalla}
\affiliation{Department of Physics and Astronomy, University College
  London, London, WC1E 6BT}

\begin{abstract}
Suggestions have been made that the microwave background observed by
COBE and WMAP and dubbed Cosmic Microwave Background (CMB) may have an 
origin within our own Galaxy or Earth.
To consider the signal that may be correlated with Earth, a correlate-by-eye
exercise was attempted by overlaying the CMB map from Wilkinson Microwave Anisotropy Probe on a topographical map of Earth.
Remarkably, several hot spots in the CMB map 
are found to be well aligned with either large cities on Earth or regions of high altitude.
To further study the correlations between Earth and CMB, we performed
a complicated cross-correlation analysis in the multipole space.
The overall correlations are detected at more than 5 $\sigma$ confidence level.
These results can be naively interpreted to suggest  that  large angular scale
fluctuations in CMB are generated on Earth by a process that traces
the altitude relative to a mean radius.   Simply extending our analysis, we suggest that cross-correlations between CMB
and any other map of a Solar system body, image of a person, or an image of an animal
will be detected at some statistical significance. It is unclear how Occam's razor
can be applied in such a situation to identify which sources are responsible for CMB fluctuations.
\end{abstract}


\maketitle

\noindent \emph{Introduction--- } Despite over three decades of intense theoretical 
and experimental studies with over a billion
dollars of research funds spent over that time period including two dedicated space-based missions and several Nobel prizes,
the origin of the Cosmic Microwave Background (CMB) is still hotly debated at occasional research conferences, 
in referred journal articles \cite{Robit,del}, 
and by members of an active blogosphere on non-standard cosmological models\footnote{Interested readers are
encouraged to Google relevant keywords}. Among non-cosmic suggestions for 
the origin of CMB is the possibility that it originates from Earth \cite{Robit}
\footnote{CMB may be described as the Ocean Microwave Background (OMB).} 
Other suggestions include the possibility that CMB, or at least
some fraction of the CMB intensity, may have a Galactic origin.
Here, we study the possibility that CMB originates from Earth by performing several
analysis involving the extent to which CMB anisotropy intensity distribution correlates with information on Earth's
surface such as the altitude. For this, we make use of both
``correlate-by-eye'' (a simple variation of well utilized ``chi-by-eye'' technique) 
as well as a more sophisticated harmonic analysis.\\

\begin{figure*}[!t]
\centerline{\psfig{file=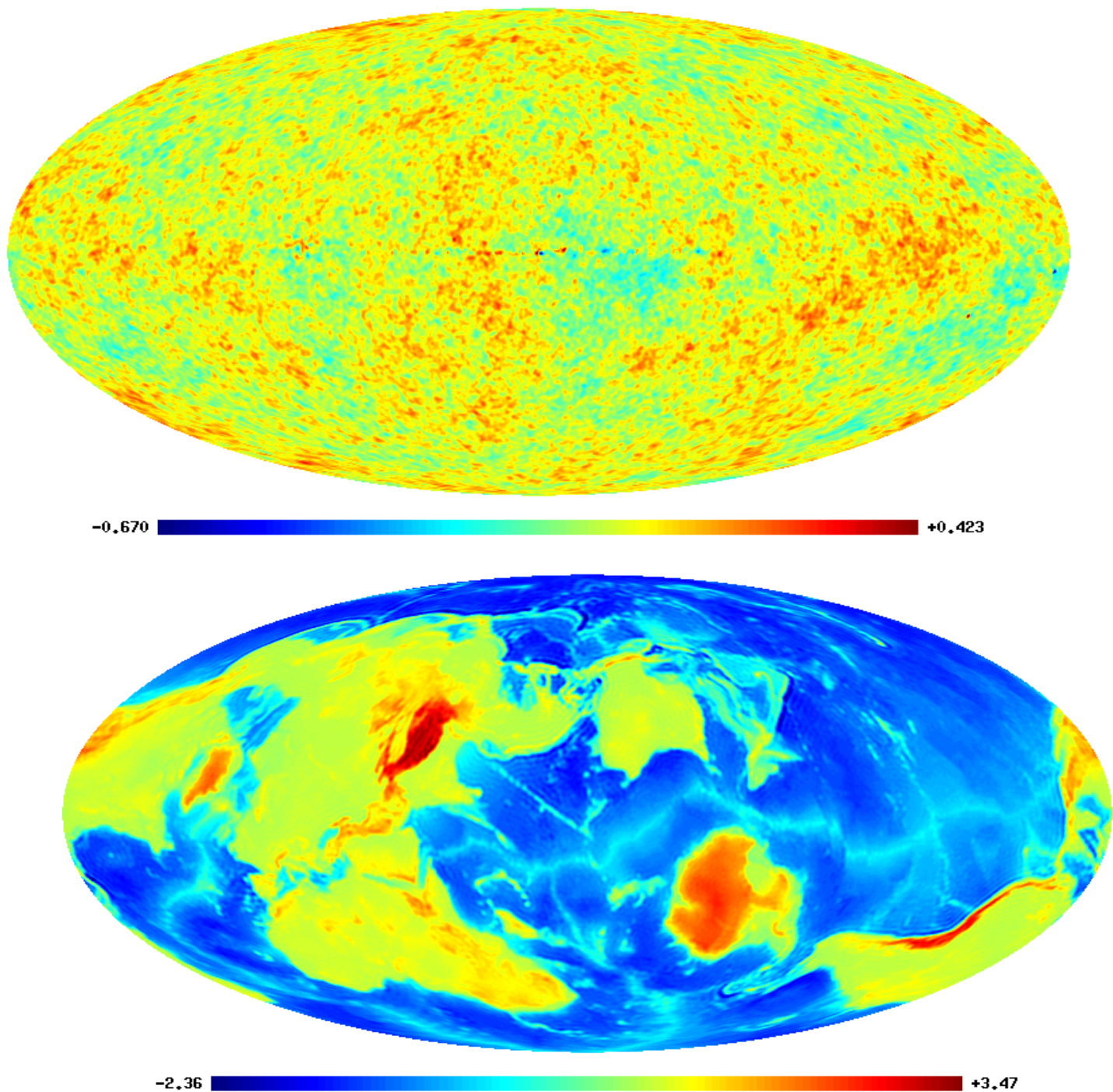,width=7.6in,angle=0}}
\caption{{\it Top:} Internal Linear Combination (ILC) map from WMAP.
{\it Bottom:} Topographical map of Earth in the Galactic coordinate system. 
A simple correlation-by-eye indicates that a significant
number of hot spots (in red) overlap within a distance of 1 deg$^2$ of major cities on Earth's map
as well as regions of high altitude (in red).
Other significant overlaps between the two maps are also clear from a large cold region (in blue)
in WMAP corresponding to the Indian Ocean.
\label{maps}}
\end{figure*}

\noindent \emph{Data--- } To perform this study, we first downloaded the high signal-to-noise Internal Linear Combination  (ILC) CMB map  produced by 
the Wilkinson Microwave Anisotropy  Probe (WMAP) team\footnote{Available from http://lambda.gsfc.nasa.gov/}.  
This map combines observations at several different
frequencies to remove Galactic foregrounds given their frequency spectra relative to the
thermal black body CMB.
We also obtained a perfect map of the Earth topography\footnote{A fits
  file of the map portable to HEALPix is available at
  http://astro.ic.ac.uk/~pdineen/earth/index.html}. These maps are
reproduced in Figure~\ref{maps} for reference. Instead of the equatorial coordinate system usually employed to display Earth, we convert the altitude map to the galactic coordinate system since the WMAP  CMB map is in that coordinate system.\\

\noindent \emph{Method--- } To extract which signatures in the CMB map are correlated with features on Earth's surface, we first performed a 
first-step ``correlate-by-eye''
study in the same manner ``chi-by-eye'' analysis are used frequently  to interpret astronomical observations.
The correlate-by-eye technique involves either overlaying the CMB map on Earth's map or Earth's map on top of the CMB map. 
We believe both approaches produce the same results so only the former method is used here.
As can be deciphered by carefully staring at Figure~\ref{maps} for about 2 minutes,
an unusually large number of well-known large cities on Earth, as well as high altitude areas,
 are found within few degrees from regions with a large density of hot spots (or red regions)
 in the WMAP map. The Indian ocean for example clearly appears as a colder region relative to the mean CMB temperature
(represented by blue). If interpreted through this overlap, we can claim that
CMB features are due to some kind of an unknown physical process on Earth that manages to radiative
as a blackbody but the correlate-by-eye analysis must be extended to consider statistical chances and
random coincidences to fully understand the significance of this result.

To access the probability that there is a significant overlap between the two maps (for example to establish
whether the boundaries of the continents are visible in the CMB map), we performed a battery of statistical tests ranging from
complex applications of the Bayes' theorem to computationally challenging and time consuming Markov Chain Monte Carlo estimates
of the likelihood function.
Due to limited space allowed for this article, we neither describe nor reproduce exact details of these statistical 
tests. However, we want the reader to know that we have carefully analyzed all possibilities\footnote{Interested parties 
can obtain this information from the first author}. Our statistical tests reveal that once certain regions are
masked out or ignored (for example, least populated regions on Earth), 
then there are indeed large correlations between the two maps. While the correlate-by-eye
and associated statistics are useful, such studies remove regions of the map that are too complex to comprehend
or analyze. To make use of all data  we must investigate this correlation further with a more complicated analysis.

\begin{figure}[!t]
\centerline{\psfig{file=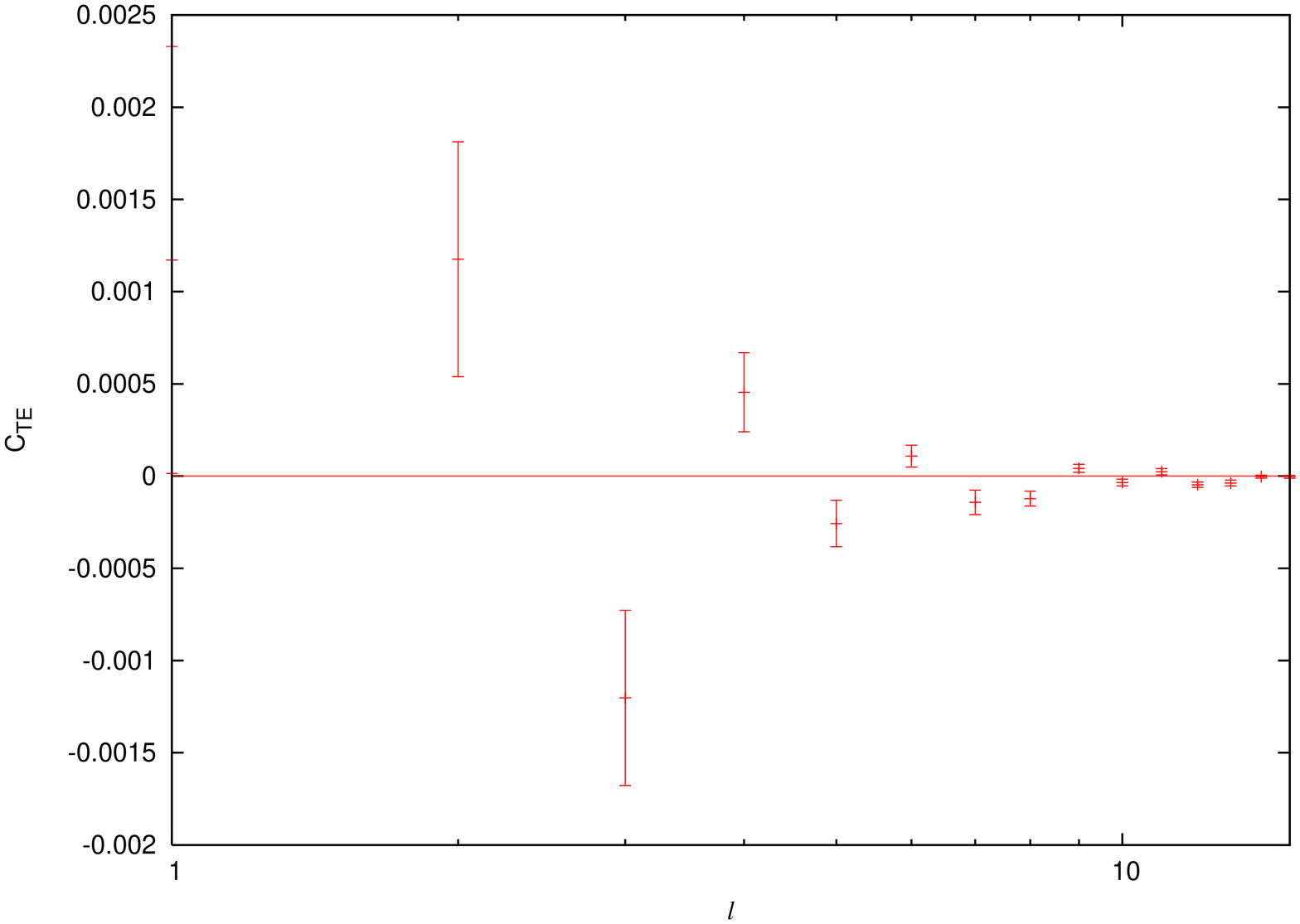,width=3.3in,angle=-90}}
\caption{The cross-correlation power spectrum between the two maps in Figure~1.
The error bars are derived with a complicated analysis in the same manner error bars derived for
cross-correlations between CMB and galaxies. The cross-correlation is detected at very high significance
and is more significant than the detection of the supposedly
cosmological ISW effect.  The line plotted at $C_{TE} = 0$ corresponds
to the null hypothesis of no correlation.\label{cross}}
\end{figure}

It is common in CMB studies, especially those studies where cosmological ISW signal is extracted by comparing to 
maps of the galaxy distribution,  to cross-correlate maps by decomposing them to multipole moments.
Here, we have implemented a similar analysis using
 HEALPix\footnote{http://healpix.jpl.nasa.gov/}. The cross-correlation involves
taking multiple moments of the two maps $a^i_{lm} = \int d{\hat{\bf n}} M_i(\hat{\bf n}) Y^*_{lm}(\hat{\bf n})$,
where $M_T(\hat{\bf n})$ is the temperature anisotropies of the CMB
(corresponding to $a^T_{lm}$) and $M_E(\hat{\bf n}))$ the Earth's altitude relative to the mean radius
(corresponding to $a^E_{lm}$) and performing $\langle a_{lm}^{\rm T} a_{lm}^{\rm *
  E}\rangle$ to establish the angular cross power spectrum. We assume
statistical error bars due to gaussian $a_{lm}$'s, which are
calculated using: 
\begin{equation}\sigma^2(C_{TE})=\frac{1}{2\ell+1}(C^2_{TE}+C_{TT}C_{EE})\end{equation}

If the cross power spectrum were to be exactly zero then CMB is independent of surface features on Earth, but if it  is
not zero, then one may be able to argue that some connection exists between the two maps even without
a theoretical model for a connection.  In fact, we believe a {\it model
  independent} detection of a cross-correlation constitutes stronger evidence of a
correlation than a {\it model dependent} detection actually based on a
model with physical intuition and/or theoretical justification.\\

\noindent \emph{Results and Discussion--- }Our results related to the cross angular power spectrum are summarized
in Figure~\ref{cross} where we find that there
is indeed an excess correlation at multipoles less than 10. These correlations behave such that for
even multipoles the two maps are positively correlated while for odd multipoles the two maps
are negatively, or more appropriately, anti-correlated. We have no physical model or a process connecting 
the CMB sky and Earth to
explain why the sign is changing between the even and odd multipoles. While we also do not have
a theoretical model to explain the full amplitude of the cross-correlation,
one can naively interpret this cross power spectrum as evidence that some of the signal in the WMAP CMB map
originates from height variations relative to the mean radius of Earth. 
As the mean radius corresponds to the mean sea level, this study may suggest
that some of the low $\ell$ power in the microwave background
fluctuations is in fact due to the OMB - if it exists.  Overall, the cross-correlation is
detected at more than 5 $\sigma$ relative to a model with no correlations between these two maps. This is remarkable given that
the cross-correlation between WMAP CMB map and the galaxy distribution is only present at 2$\sigma$ to 3$\sigma$ level
\cite{Ras}. 

The correlated signal from WMAP to Earth is only present at
scales corresponding to a few tens of degrees and more suggesting that some fraction of large angular scale
anisotropies may not be cosmic. The small angular scale anisotropies, such as the acoustic peaks
are not correlated with Earth and if CMB does not originate from the early Universe then
there is still a large missing fraction of CMB that must be accounted for. 
To study the true origin of the CMB intensity distribution, we encourage further cross-correlation studies of CMB map
against other maps that can be projected on the sky using spherical coordinates. Possibilities for this
include any pictures that one might be able to find, such as Moon, Mars or even a picture of your favorite
actor or actress. Our studies show that, despite having a strong theoretical motivation why such cross-correlations must be
performed, it may be possible that detection (or non-detection) of a
cross-correlation signal may actually be a statistical fluke.
This observation then clearly leads to a complex situation. Since any map is likely to be correlated with CMB at some
significance, should we always assume that some connection exists between CMB and that map or the sources or tracer field
represented by that map?  We leave this question open for the
scientific community to debate.

Complicated studies using multipole vector statistics that are due to Maxwell have shown that there is
an axis of evil on the CMB map \cite{Land}. This axis is present through alignments of low multipole 
moments of the CMB map and there is no good explanation for this axis within the standard cosmological model \cite{Hut}.
Another axis of evil is also present on Earth\footnote{George W, Bush, State of the Union Address, January 29, 2002}.
We considered the possibility that the Earth's axis of evil aligns with the CMB axis of evil. Correlate-by-eye analysis
presents no significant common features between the member states of Earth's axis of evil\footnote{see http://en.wikipedia.org/wiki/Axis-of-evil} and CMB axis of evil. Since we only performed correlate-by-eye study
on the alignment of the two evil axes and did not make use of Maxwell's multipole vectors and other
complex statistics, we cannot confidently state that the two evil axes are independent. A careful study
must be conducted and is strongly encouraged.

Finally, we wish to comment on an existing suggestion in the literature  that there are hidden messages from the creator 
in the WMAP data since it can be thought of as a billboard visible throughout
the Universe so a message is likely to be encoded within the intensity fluctuations \cite{Hsu}. 
In the extreme case that this message is in fact an image of the creator
herself hidden within the cosmic noise, we suggest that it may be possible to
establish this image through a large number of cross-correlations of input images (of people, animals, spirits, or
combinations)  that are used in  a likelihood analysis. The image, or an image reconstructed from an ensemble,
that maximizes the likelihood in such a comparison can be considered to be the best reconstructed image of the creator.
Unfortunately, the information content of the CMB is limited to be
about 300 bits \cite{Scott} so the image is likely to be extremely low resolution. Still, we encourage a search for such an image
especially to study  the possibility that  the creator  is human. In fact, we would not be surprised if a spherically projected
image of a famous celebrity correlates with WMAP at some high significance. Thus, it is left to the
reader to establish using a complicated theoretical argument if that celebrity is the creator whose image is then hidden in WMAP.

\begin{acknowledgments}
\emph{Acknowledgments--- } We acknowledge all participants of the recent {\it Outstanding Questions for the Standard Cosmological
Model} conference at Imperial College, London for the breadth of presentations and discussions on the
status of our ``standard cosmological model'' and alternative
interpretations and suggestions.
We also thank the pub down the Gloucester Road for not closing
exactly at the time it should have closed to give us an extra
half-hour to come up with the basic idea for this paper. AR would also
particularly like to thank a
senior member of the cosmology community for suggesting she cross-correlate {\it ``anything she could get her hands on''}.
This research work did not use any federal funds.  Any important outstanding
questions related to this article and its implications should be directed to the first author.
\end{acknowledgments}

\end{document}